\documentstyle[11pt,paspconf,164page,psfig]{article}

\begin{document}

\title{Sgr A* and Company - Multiwavelength observations of Sgr A* and
VLA search of ``Sgr A*'s'' in LINERs\footnote{Talk presented at IAU Coll.~164, 
A.~Zensus, G.~Taylor, \& J.~Wrobel (eds.), PASP Conf.~Proc.}}
\label{falck}

% running head stuff
\markboth{Falcke et al.}{Sgr A* and Co.}

\author{
H. Falcke\altaffilmark{1,2}, 
W.M. Goss\altaffilmark{3}, 
L.C. Ho\altaffilmark{4}, 
H. Matsuo\altaffilmark{5}, 
P. Teuben\altaffilmark{1},
A.S. Wilson\altaffilmark{1}, 
J.-H. Zhao\altaffilmark{4}, 
R. Zylka\altaffilmark{2,6}
}

\altaffiltext{1}{Dept. Astronomy, University of Maryland, College Park, MD
20742-2421, USA}
\altaffiltext{2}{Max-Planck-Institut f\"ur Radioastronomie,
Auf den H\"ugel 69, D-53121 Bonn, Germany.}
\altaffiltext{3}{NRAO Socorro}
\altaffiltext{4}{Harvard/CfA}
\altaffiltext{5}{NRO/Japan\quad$^6$ ITA Heidelberg}

\begin{abstract}
We report first results from a multiwavelength campaign to measure the
simultaneous spectrum of Sgr A* from cm to mm wavelengths. The
observations confirm that the previously detected submm-excess is not
due to variability; the presence of an ultracompact component with a
size of a few Schwarzschild radii is inferred.  In a VLA survey of
LINER galaxies, we found Sgr A*-like nuclei in one quarter of the
galaxies searched, suggesting a link between those low-power AGN and
the Galactic Center.
\end{abstract}

\section{Introduction}
The closest compact, flat spectrum radio core in the center of a
galaxy is Sgr A* in the Galactic Center (GC). NIR observations of the
GC have convincingly demonstrated the presence of a dark mass of
$2.5\cdot10^6 M_{\odot}$ (Eckart \& Genzel 1996) which is most likely
due to a black hole associated with Sgr A*. Here we present the
results of two observational campaigns that may shed further light on
the nature of Sgr A* and of flat spectrum radio cores in galactic nuclei in
general.

%\index{Sgr A*}
%\index{compact radio cores}

\section{Multiwavelength campaign for Sgr A*}
Early mm- and submm-observations of Sgr A* suggested the presence of a
submm-excess in the spectrum of Sgr A* (Zylka et al.  1992). Since Sgr
A* can be variable it was, however, not clear, whether this excess was
due to non-simultaneous measurements, a systematic error, or an
intrinsic up-turn of the spectrum. Nevertheless, so far the excess at
submm-wavelengths has persisted in non-simultaneous measurements by
various other groups (see Morris \& Sera\-byn, p. 685). As
a further step, we have now performed a campaign to measure
quasi-simultaneously the spectrum of Sgr~A* to exclude the effects of
variability on the broad-band radio spectrum.
%\index{spectrum!Sgr A*}

\begin{figure}[th]
\centerline{
\psfig{figure=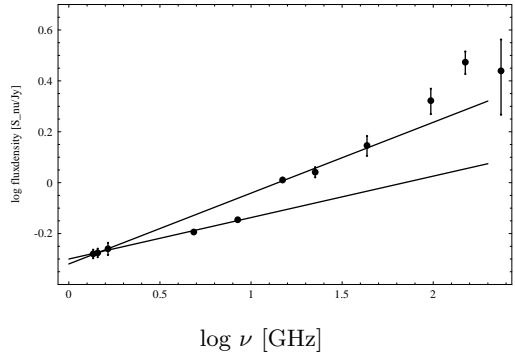,width=0.5\textwidth,bbllx=3.4cm,bburx=18.3cm,bblly=2.5cm,bbury=25.4cm,angle=-90}
}
\centerline{\footnotesize log $\nu$ [GHz]}
\caption{Spectrum of Sgr A* averaged over all telescopes at each
wavelength for October 25-27, 1996.\label{falckf1}}
\end{figure}

The observation were carried out at four different telescopes---VLA
A-Array (20, 6, 3.6, 2, 1.3, \& 0.7 cm), IRAM (3, 2, \& 1.3 mm), BIMA
C-Array (3mm), and Nobeyama 45m (3 \& 2 mm)---on three consecutive
days on October 25/26/27, 1996. The data were reduced by the
individual groups following standard procedures. For BIMA we did not
use the longer baselines because of coherence problems; the IRAM 1.3mm
observations were affected by weather.

The resulting spectrum is shown in Fig.~1 where we have averaged the
fluxes at each wavelength from all days and telescopes. The
VLA spectrum shows a marked break at 10 GHz. Below the break it can be
described as a powerlaw $S_{\nu}\propto\nu^{\alpha}$ with
$\alpha=0.16$ and above the break as a powerlaw with
$\alpha=0.28$. The combined 3 and 2mm fluxes seem to be significantly
above the extrapolation from the VLA fluxes. As discussed in Falcke
(1996) such an excess, if due to self-absorption, indicates an
ultra-compact region of $\sim2-3$ Schwarzschild radii. This region
could in principle be resolved by future, global (sub)mm-VLBI
experiments and thus one could directly probe the black hole nature of
Sgr~A*.

\section{VLA survey of nearby LINER galaxies}
To see whether Sgr A* is really unique, we have surveyed a sample of
48 nearby LINER galaxies, selected from Ho et al.~(1995), with the VLA
in A-configuration at 2cm. We found that a quarter (mostly spirals) of
the galaxies surveyed showed a compact, flat-spectrum core above a 5
$\sigma$ detection limit of $\sim1$ mJy. Hence, we conclude that a
nucleus like Sgr~A* is not a unique feature of our galaxies, but can
be found in other nearby galaxies, especially in those with signs of
optical nuclear activity. Since the cores we found show a good
radio/H$\alpha$ correlation, it is very likely that they are
indeed directly associated with the nuclear engine.
%\index{jets in!LINER galaxies}

%\begin{figure}[h]
%\centerline{
%\psfig{figure=corr2.ps,width=0.3\textwidth,bbllx=2.5cm,bburx=19.2cm,bblly=6.1cm,bbury=21.8cm}
%}
%\centerline{\footnotesize $L_{\rm disk}$ [erg/sec]}
%\caption{Radio/optical correlation for flat-spectrum radio nuclei as
%given in Falcke \& Biermann (1996) (log $\nu L_{5{\rm GHz}}$ vs. log $L_{\rm
%disk}=\log{L_{{\rm H}\alpha}^{\rm narrow}+2.1}$) --- the radio cores in LINER galaxies we
%found are given as big, filled dots in the range $42<L_{\rm disk}<44$.
%\label{corr}}
%\end{figure}

\footnotesize\acknowledgments

Observations of Sgr A* were performed by MG \& J.-H.Z (VLA), RZ
(IRAM), PT \& HF (BIMA), HM (Nobeyama). The LINER survey was prepared
and performed by HF, LH, \& AW. This work was in part supported by
NASA and NSF (NAGW-3268, NAG8-1027, AST9529190).
The National Radio Astronomy Observatory is a facility
of the National Science Foundation, operated under a cooperative agreement 
by Associated Universities, Inc.

\end{document}